\documentclass[12pt,preprint]{aastex}

\newcommand{\rxte}{{\it RXTE}}
\newcommand{\swift}{{\it Swift}}

\newcommand{\etal}{et al.}

\begin{document}

\title{Discovery of Eclipses from the Accreting Millisecond X-ray 
Pulsar SWIFT J1749.4$-$2807}
\author{C. B. Markwardt,\altaffilmark{1,2}
        T. E. Strohmayer \altaffilmark{3}}

\altaffiltext{1}{Department of Astronomy, University of Maryland,
        College Park, MD 20742; Craig.Markwardt@nasa.gov}
\altaffiltext{2}{Astroparticle Physics Laboratory,
        Mail Code 661, NASA Goddard Space Flight Center, Greenbelt, MD 20771}
\altaffiltext{3}{X-ray Astrophysics Laboratory,
        Mail Code 662, NASA Goddard Space Flight Center, Greenbelt, MD 20771}

\begin{abstract}
We report the discovery of X-ray eclipses in the recently discovered
accreting millisecond X-ray pulsar Swift J1749.4$-$2807. This is the
first detection of X-ray eclipses in a system of this type and should
enable a precise neutron star mass measurement once the companion star
is identified and studied.  We present a combined pulse and eclipse
timing solution that enables tight constraints on the orbital
parameters and inclination and shows that the companion mass is in the
range $0.6 - 0.8 M_{\odot}$ for a likely range of neutron star masses,
and that it is larger than a main sequence star of the same mass. We
observed two individual eclipse egresses and a single ingress. Our
timing model shows that the eclipse features are symmetric about the
time of $90^{\circ}$ longitude from the ascending node, as expected.
Our eclipse timing solution gives an eclipse duration (from the
mid-points of ingress to egress) of $2172\pm13$ s. This represents
6.85\% of the 8.82 hr orbital period.  This system also presents a
potential measurement of ``Shapiro'' delay due to General Relativity;
through this technique alone, we set an upper limit to the companion
mass of 2.2 $M_\sun$.

\end{abstract}


\section{Introduction}

The neutron star mass (and radius) distribution has important
implications for our understanding of the equation of state (EOS) of
ultra-dense matter. The EOS sets directly a number of potentially
observable quantities, including the maximum mass and spin frequency
of a neutron star (Lattimer \& Prakash 2001). Moreover, measurements
of neutron star radii would strongly constrain the nuclear symmetry
energy (Steiner \etal\ 2010).  At present there remain serious gaps in
our understanding of the mass distribution of neutron stars,
particularly as regards their upper mass limit. Where neutron star
masses are known with precision it is in most cases for relatively
young, binary radio pulsars in which one or more relativistic, Post 
Keplerian (PK) parameters can be directly measured (Thorsett \& 
Chakrabarty 1999). Two famous examples of such systems include the 
original Hulse - Taylor binary pulsar PSR 1913+16 (Taylor \& Weisberg 1989), 
and the double pulsar PSR J0737$-$3039 (Kramer et al.  2006). These neutron 
stars have masses that appear to cluster near or below the ``canonical'' 
value of 1.4 $M_{\odot}$, and hence do not strongly constrain the maximum 
mass nor the EOS.

Studies of older, accreting neutron star systems hold significant
promise for directly probing the high end of the mass
distribution because it is likely these objects have accreted significant 
mass during their evolutionary history.  
Pulsar timing studies of binary millisecond radio
pulsars have recently led to strong indications for higher mass
neutron stars. For example, Freire et al. (2009) 
present timing measurements of PSR J1903+0327 that indicate its neutron 
star has a mass of $1.67\pm0.01 M_{\odot}$ (also Champion et al. 2008). 
A related population of 
sources which could provide important mass constraints are 
accreting millisecond X-ray pulsars (AMXPs). Additionally, 
these systems also produce X-ray emission directly from the neutron star 
surface in the form of thermonuclear X-ray bursts, and in principle
can also be used to constrain neutron star radii, something that 
cannot be done for the radio pulsar systems unless spin orbit coupling
can be directly observed (Lattimer \& Schutz 2005). Currently twelve
of these systems are known, with the first, \object[SAX J1808.4-3658]{SAX J1808.4$-$3658}, having
been discovered in 1998 (Wijnands \& van der Klis 1998).  A subset of
the AMXPs are ultracompact binaries, with orbital periods less than an
hour and have very low mass ($<< 0.1 M_{\odot}$), degenerate dwarf
companions.  An example of a system of this type is \object[XTE J1751-305]{XTE J1751$-$305}
(Markwardt et al. 2002). The remainder appear to more closely resemble
the ``classical'' accreting, neutron star low mass X-ray binaries (LMXBs), 
with orbital periods of ~hours and low mass main sequence companions (such
as SAX J1808.4$-$3658).

Several factors currently limit the use of these systems to obtain
precise neutron star mass estimates. First, the population is largely
associated with the Galactic bulge. Therefore the resulting high
optical extinction makes detection of companions
challenging. Second, their binary inclinations have, to date,
not been strongly constrained, as would be possible, for example, with
the detection and timing of X-ray eclipses. Indeed, observations of
eclipsing AMXPs would enable the binary inclination to be tightly
constrained and, combined with precision X-ray pulse timing and
spectroscopic measurements of the companion, could lead to precise
mass constraints for their neutron stars. Finally, it has not yet been
possible to measure any PK parameters (such as the Shapiro delay) 
in these systems.

In this Letter we report the first detection of X-ray eclipses from an
AMXP, in the system \object[SWIFT J1749.4-2807]{SWIFT J1749.4$-$2807}
(hereafter J1749).  J1749 was first discovered in 2006 based on the detection 
with \swift/BAT of a thermonuclear X-ray burst (Schady et al. 2006).  
Follow-up observations with \swift\ set an upper limit to the source 
distance of $6.7\pm1.3$ kpc (Wijnands et al. 2009).
The source was detected again in 2010 by {\it INTEGRAL}, exhibiting both 
persistent and bursting activity (Pavan et al. 2010; Chenevez et al. 2010).
Altamirano et al. (2010a), using \rxte\ observations
during the recent outburst of this object, found that it harbors a 518
Hz accreting millisecond pulsar, with strong power also evident at the
1036 Hz first harmonic (Bozzo et al. 2010).  Analysis of additional \rxte\
observations first by Belloni et al. (2010) and then by
Strohmayer \& Markwardt (2010) found an 8.82 hr circular
orbit, and eclipses centered at the orbital phase of superior
conjunction of the neutron star (Markwardt et al. 2010).  
Here we present a timing analysis of the pulsar and its eclipses, 
and derive constraints on the properties of the binary components.
Our goal in this work is to establish the orbital solution from pulse
timing to a sufficient degree of precision to place the eclipse observations
in context.  We leave a detailed characterization of the pulse profiles 
and timing noise to other work (Altamirano et al. 2010b; Ferrigno et al. 2010).

\section{Observations}

The \rxte\ PCA instrument (Jahoda et al. 2006) observed J1749 between 
2010-04-14 and 2010-04-20 (UTC) for a combined total exposure of 45 ks
(observation ID \dataset[ADS/Sa.RXTE#P/95085-09]{95085-09}).  PCA
observations were performed with data mode ``E\_125us\_64M\_0\_1s,''
which reports individual photon times with a resolution of
122 $\mu$s.  We corrected event arrival times to the solar system
barycenter using the \swift/XRT source position (Yang et al. 2010;
Table \ref{tab-orbit}).

Over the course of the \rxte\ observing campaign, the
2--10 keV flux of the source decayed from a peak of $\sim$9 mCrab, to
an approximate quiescent level.  J1749 is located 0.32$^\circ$ from
the galactic plane and 1.17$^\circ$ from the galactic center.  The PCA
field of view is approximately 1.1$^\circ$ (radius to zero-response),
so galactic diffuse emission and point sources in the field of view
also contribute to the total count rate.  We determined background
count rate in the 2--30 keV band to be 17.5 ct/s/PCU for PCUs 2--4 and
21.4 ct/s/PCU for PCUs 0 and 1, based on data taken after 2010-04-20,
when the source had reached near-quiescence (also confirmed by \swift/XRT 
imaging observations; Yang et al. 2010).

Upon visual inspection, there are sharp ``dip-like'' features in the
X-ray light curve (Fig. \ref{fig:eclipse}.  Here, we based our analysis on
data taken by the PCA in ``Standard2'' mode, which has 16 second time bins
and 129 bins of energy resolution.  It was verified that the
dips were not due to satellite or PCA instrumental effects such as
earth occultations or instrument turn-ons or -offs.  Furthermore, the
dips coincided with the time when eclipses would be expected from the
timing solution (see below).  Therefore we concluded that it was
likely we were seeing X-ray eclipse features produced by the companion star.
We did not detect any other orbit-related modulations.

\section{Data Reduction}

As reported previously (Altamirano et al. 2010a; Bozzo et al. 2010),
the pulsed signal is detectable at both the fundamental frequency of
518 Hz and the first harmonic at 1036 Hz.  The pulsed signal of the
fundamental was intermittent, whereas we have found the first harmonic
to always provide a reasonably strong signal.  Therefore, we proceeded
with a pulse timing analysis using the first harmonic.  We used the
$Z^2$ statistic (or Rayleigh's statistic; Buccheri et al. 1983; also
described in Markwardt et al. 2002).  Signal from each detected count
in the 2--30 keV band (PCA channels 5--80) is added coherently using a
trial pulse and orbital model.  The pulse model includes a constant
pulse frequency, $f_o$, and a frequency derivative term, $\dot{f}$, if
desired.  The orbital model is derived from the \verb|ELL1| model
(Wex, unpublished; Lange et al. 2001) of the radio pulsar timing
software ``TEMPO 2'' (Hobbs et al. 2006).  This model is advantageous
for low-eccentricity binary systems where it may be difficult to
disentangle the classical eccentricity and longitude-of-periastron
orbital parameters.  The parameters are adjusted until the maximum
$Z^2$ signal is obtained. By examining smaller segments of data, it is
possible to construct a timing pseudo-residual using the phases of the
cosine and sine terms in the $Z^2$, which describes effective time of
arrival of the first harmonic pulse compared with the model.


The apparent eclipse features in the 2--30 keV X-ray light curve were
fitted by a simple eclipse profile model.  The model consists of a
constant flux level in-eclipse and out-of-eclipse, and a linear
transition between the two flux levels at a specified time.  For the
purposes of this analysis, we performed an individual analysis of the
eclipse features in order to derive their times of arrival
independently.  Eclipse occurences are
expected to be centered at the epoch of pulsar
longitude 90$^\circ$, $T90$, as measured from the ascending node. We
also fitted a model to all the data jointly where the ingress/egress
time, as measured from 
$T90$, was a single parameter.  Such an analysis gives a single
weighted estimate of the eclipse duration, assuming that it is
centered on $T90$.  Background subtraction was performed as described
above.

\section{Results}

Using the pulse timing techniques described above, we found a single 
coherent pulse timing solution. Timing residuals before 2010-04-19 are less
than 130 $\mu$s, equivalent to $<13\%$ of the harmonics's period; after
that date, the statistical errors become larger.  The
best pulse timing solution is shown in Table \ref{tab-orbit}.  The
harmonic pulse semi-amplitude ranges from about 8\% to as high as 30\%
of the persistent flux during the outburst.  The standard deviation of
all residuals is 90 $\mu$s, with evidence of red timing noise in
the residuals, so parameter uncertainties have been scaled upward by
$\sqrt{\chi^2/\nu}$ in order to present a more conservative picture of
the uncertainties.  We attempted to add a pulse
frequency derivative and eccentricity terms to the model.  In the case
of eccentricity, we report a marginal value, corresponding to
2.8$\sigma$.  Given the presence of red noise, it appears likely that this
value is not significant.  For the frequency derivative term, we set
only an upper limit.  Both of these terms were set to zero in
subsequent analysis.  The results presented in the table are slightly
different than those reported by Strohmayer \& Markwardt (2010),
because of the inclusion of more data in this work.

\begin{figure}[h]
\epsscale{0.80}
\plotone{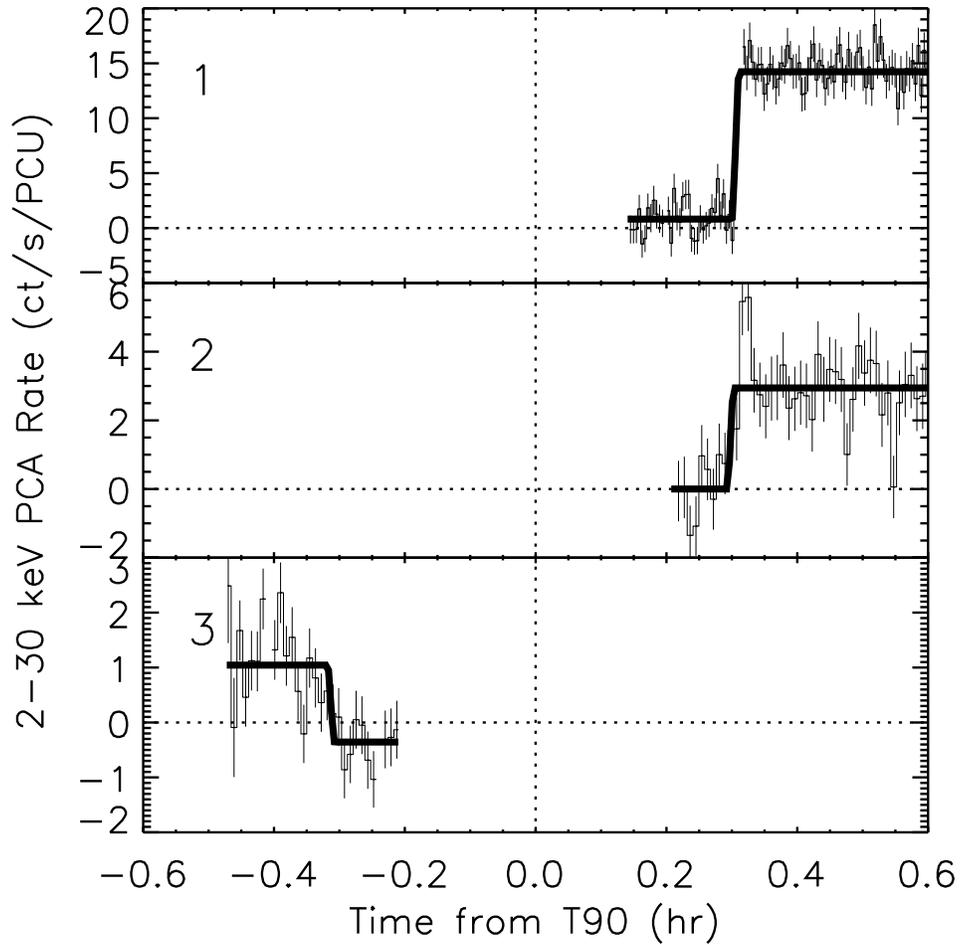}
\figcaption {X-ray light curve segments of SWIFT J1749.4$-$2807, showing 
eclipse features. Each panel is centered on the nearest epoch of $T90$, where
the pulsar is at orbital longitude $90^\circ$, and labeled by number
according to Table \ref{tab-eclipse}.  The best-fit individual 
eclipse models are also shown (thick black line).  The light curve
bin sizes are 16 s for eclipse ``1'' and 32 s for ``2'' and ``3.''
\label{fig:eclipse}}
\end{figure}

The eclipse features are shown in Figure \ref{fig:eclipse}, including
the best-fit models.  The features plotted in Figure \ref{fig:eclipse}
represent all light curve points which pass within $\sim$20 min of
$T90$.
The first eclipse egress, labeled ``1,'' is most
strongly constraining, and the other two ingress/egress features are
weaker and less constraining.  We also
verified that the pulsations were effectively eclipsed during those phases.
The results of
the individual light curve fits are shown in Table \ref{tab-eclipse}.
Those results are consistent with a single ingress/egress time,
approximately 1100 s equidistant from $T90$, and thus we are justified
in fitting the light curves jointly with a single eclipse-duration
parameter.  The results of that joint fit are shown in Table
\ref{tab-orbit}, as 
the full eclipse duration from ingress to egress mid-points.
We note that the eclipse duration reported here is somewhat different
than that reported in Markwardt et al. (2010); that report differed
due to a numerical error which was corrected in this work.

The ingress/egress duration is constrained primarily by eclipse number
``1.''  The reported egress duration of 28$\pm$14 s is consistent with
zero at the $\sim 2.5\sigma$ significance level.  Thus, we consider
this solution to be only a 95\% confidence upper limit of 55 s.

During eclipse ``1,'' we find a residual count rate of $0.8\pm0.2$
ct/s/PCU, or  $5.7\%\pm1.5\%$ of the persistent flux level.  The fitted
count rates during eclipses ``2'' and ``3'' are only 95\% confidence upper
limits of 20\% and 30\%, respectively, which are consistent with
eclipse ``1.''

\section{Interpretation}

J1749 is the first AMXP to manifest
X-ray eclipses.  The measured eclipse ingress and egress epochs are
symmetric about $T90$, as expected based on simple orbital geometry.
The presence of eclipses allows us to constrain the orbital parameters
much more closely than possible by pulse timing alone.  Our combined
pulse timing and eclipse solution is shown in Figure
\ref{fig:orbgrid}.

Here we assume that the companion star fills its Roche lobe (which
leads to a constraint on the approximate density of the companion
star; Eggleton 1983).  This assumption is reasonable because the
system is similar to other low-mass X-ray binary systems driven by
disk accretion.  The minimum companion mass is also constrained by the
measured mass function, $f_x$.  The mass function for binary systems
is defined as,
\begin{equation}
f_x = {4\pi^2 (a_x \sin i)^3 \over G P_b^2} = {(M_c \sin i)^3 \over (M_x + M_c)^2}
\end{equation}
where $G$ is the gravitational constant and other variables are
defined in Table \ref{tab-orbit}.  For the special case of an edge-on
system, $i=90^\circ$, this defines a minimum companion mass $M_c$ for
a given neutron star mass $M_x$.  The resulting curve is shown in Figure
\ref{fig:orbgrid} (``P'' line), for an example neutron star
mass of 1.4 $M_\sun$.

An eclipse duration measurement essentially determines the solid angle 
subtended by the companion as seen by the neutron star (``E'' line in
Figure \ref{fig:orbgrid}).  The actual constraint is given by (e.g. Chakrabarty et al. 1993),
\begin{equation}
\cos^2 i + \sin^2 i \times \sin^2 (\pi T_{\rm ecl}/P_b) = \sin^2(\theta/2)
\end{equation}
where $\theta$ is the angular diameter of the star as seen by the
neutron star, $R_c = a_{\rm tot} \sin(\theta/2)$, and $a_{\rm tot}$ is
the total binary separation.  For the same example neutron star mass,
this curve is shown in Figure \ref{fig:orbgrid} (``E'' line).

The constraint lines intersect at a single point for a given neutron
star mass.  Assuming that the neutron star mass lies somewhere in the
range 1.4--2.2 $M_\sun$, the joint solution produces a continuous
curve shown in Figure \ref{fig:orbgrid} (thin black line).  These joint
constraints on the companion mass, radius and inclination are also
shown in Table \ref{tab-orbit}.

For this system's estimated companion mass $M_c \simeq 0.7 M_\sun$ and nearly
edge-on inclination, the
effect of a Shapiro-like delay is appreciable (Shapiro et
al. 1971).  According to General Relativity, photons from the pulsar will
experience an additional delay due to the gravitational potential of
the companion star, with magnitude given by,
\begin{equation}
\Delta{t_s} = -2GM_c/c^2 \log(1 - \sin i \sin\phi) < 21\ \mu{\rm s} 
   \left({M_c\over 0.7 M_\sun}\right)
\end{equation}
where $\phi$ is the orbital longitude, and we used $i=77^\circ$.
Although the maximum effect is expected to occur at the center of
eclipse when pulsations are not visible, $\sim 90\%$ of the effect is
present near but outside of occultation.  In principle, a 21 $\mu$s
delay is within the grasp of \rxte's absolute timing uncertainty (Jahoda et al. 2006).

Figure \ref{fig:shapfit} (top) shows the timing residuals folded at
the orbital period.  There are no strong trends in this
representation, but significant red noise in the residuals causes the
plot to have a choppy appearance.  We attempted to remove this noise
by fitting a spline function with control points on days April 15, 16,
17, 18, and 20.  The control points were set near midnight on those
days, and rounded to the nearest point of longitude = $270^\circ$,
where the Shapiro effect is expected to be minimum.  We argue that by
fitting a spline with $\sim$daily variation, we will remove the long
term trends while leaving any more compact signal near $T90$
undisturbed.

The bottom panel of Figure \ref{fig:shapfit} shows the same residuals
after the spline trend has been removed.  The $\chi^2$ improves from
225 to 70.2 (for 59 degrees of freedom), showing the benefit of this
procedure.  Adding a Shapiro term reduces the $\chi^2$ to 67.7, which
is not significant (an $F$-ratio test would demand a reduction of
$\sim$10 to achieve 99\% confidence).  Thus, the data do not demand
the presence of the Shapiro effect.  However, assuming it is present
we set an upper limit
to the companion mass of $M_c < 2.2 M_\sun$ (95\% confidence).

\section{Conclusions}

The simultaneous presence of X-ray pulsations and X-ray eclipses have
allowed us to constrain the companion of the system with unprecedented
accuracy.  The companion mass of $\sim$0.7 $M_\sun$ is rather large
compared to most of the other AMXP
companions.  In contrast to the ``ultra-compact'' systems such as XTE
J1751$-$305, whose companion mass is a reasonable multiple of Jupiter's
mass (Markwardt et al. 2002), J1749 has a mass more comparable to the
Sun.  

\begin{figure}[h]
\epsscale{0.80}
\plotone{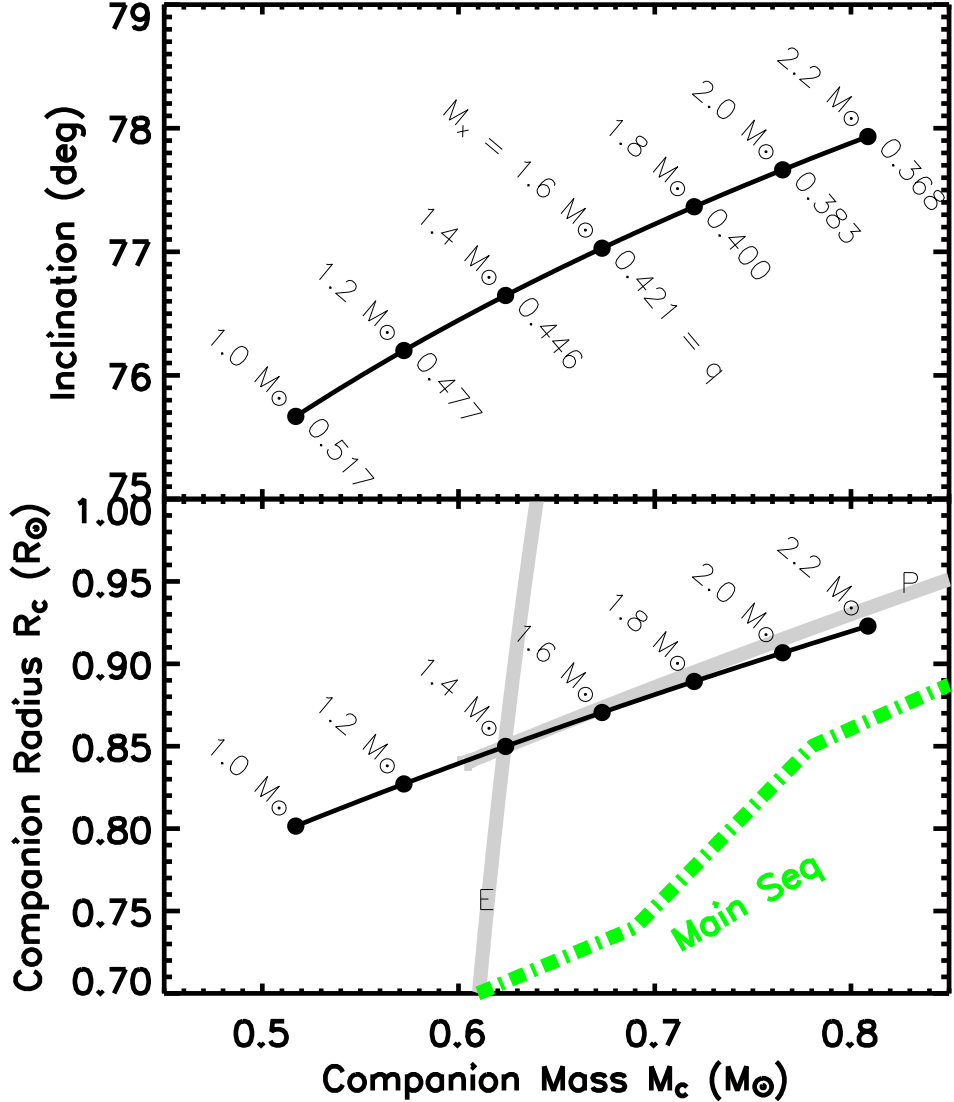} \figcaption {Joint eclipse and pulse timing solution
for J1749, assuming a Roche lobe filling companion star, for both the
binary inclination (top) and companion radius (bottom).  
The solid black lines show the allowed solutions for the measured parameters 
of J1749, for a range of neutron star masses, $M_x$, and binary mass ratios,
$q = M_c/M_x$, as indicated.
The curve for main sequence stars from Zombeck (1982) is also shown 
(green dot-dash line). The bottom
panel shows an example of how the constraints from pulse timing
and the Roche lobe constraint
(labeled ``P''); and eclipse duration (labeled ``E'')
intersect at a single point for a 1.4 $M_\sun$ neutron star.  
The actual uncertainties are thinner than the plotted lines.
\label{fig:orbgrid}}
\end{figure}

\begin{figure}[h]
\epsscale{0.80}
\plotone{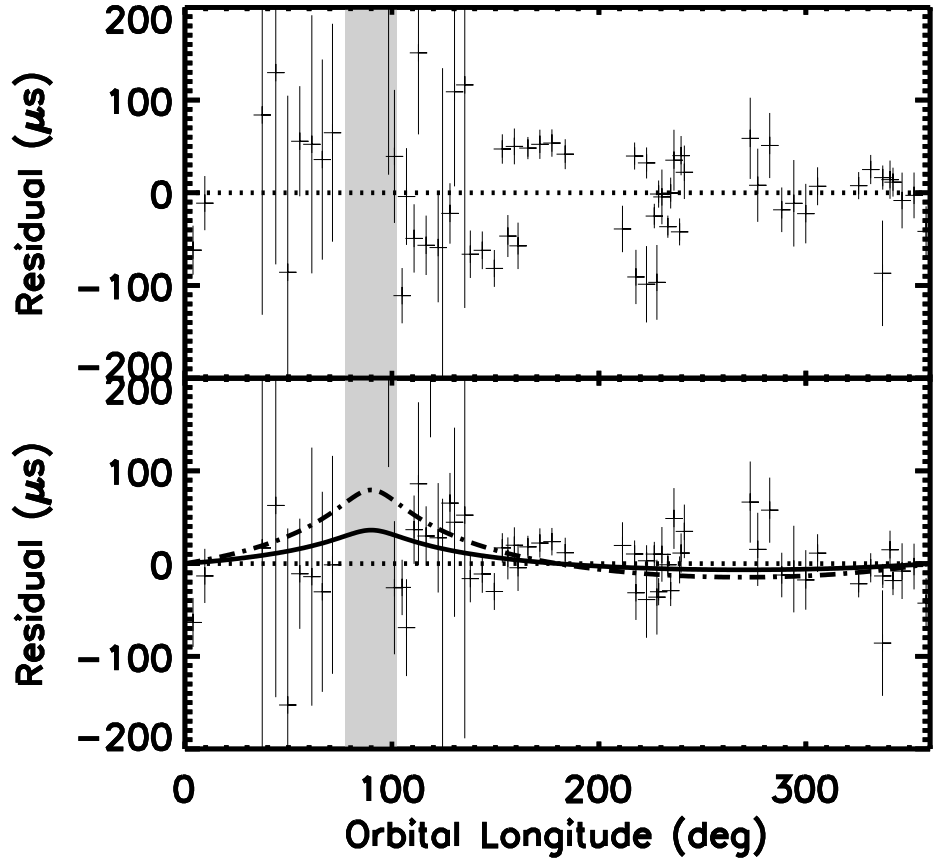} \figcaption {Search for relativistic Shapiro-like
delay in J1749.  The top panel shows the residuals as a function of
orbital phase after the orbital fit.  The eclipsed portion is indicated
by the shaded gray region.  The bottom panel shows the
same residuals after subtracting a smooth spline function (see text).
The effect of a Shapiro-like delay due to a companion mass of 1
$M_\sun$ (solid) and 2.2 $M_\sun$ (dot-dash) are shown.
\label{fig:shapfit}}
\end{figure}

The companion of J1749 appears to be larger than a typical main
sequence star, as obtained from Zombeck (1982; see Figure
\ref{fig:orbgrid}), but also from more modern studies (Chabrier et
al. 2007).  For a neutron star of 1.4 $M_\sun$, the companion is about
20\% larger than expected from the main sequence track.  We can
interpret this as ``bloating'' of the companion due to X-ray
irradiation by the neutron star which will cause the outer layers of
the star to expand (Bildsten \& Chakrabarty 2001).

The eclipse timing presented here places relatively tight constraints
on the structure of the mass donor.  Theoretical calculations indicate
that the donor in a mass transfer binary can be perturbed away from a
main sequence spectral type -- mass relationship (see Beuermann et
al. 1998; Kolb, King \& Baraffe 2001).  Beuermann et al.  (1998) compare
observational results for the donors in CVs with results of
theoretical calculations, and show that systems with orbital periods
longer than about 6 hours span a range of spectral type, whereas short
period systems (less than 2 hr) behave more like solar-abundance,
main-sequence stars.  In related work, Kolb, King \& Baraffe (2001)
show that for a given measured spectral type a range of masses is
possible based on the evolutionary status of the donor and its mass
transfer history.  The limits from their Table 1, when compared with
our companion mass estimates, would
suggest a spectral type for the secondary of J1749 from about K1 to
K3.  This range is also consistent with the results of Beuermann et
al. (1998) for an $\approx 9$ hr binary period (see their Fig. 5).

There is only one orbital variable left in the system to be known.  If
the orbital inclination or companion mass or radius could be
determined, then the neutron star mass could also be determined. In
general, spectroscopic measurements of the companion to determine its
radial velocity profile would provide the mass ratio and thus fully
constrain the system components, including the neutron star mass.  As
noted in the introduction, systems like J1749 provide a unique
opportunity to probe the high mass end of the neutron star mass
distribution.  Before this can be accomplished, an IR/optical
counterpart must be identified.  Unfortunately, high resolution
imaging {\it Chandra}
observations performed shortly after the current outburst failed to
detect an X-ray source (Chakrabarty et al. 2010), so IR/optical
observations will be difficult.  Nevertheless, spectroscopy should be
a focus of future observations when the source reappears.

We believe this is the first time that realistic limits have been set
for the Shapiro delay effect at X-ray wavelengths, for a system
outside of our solar system.  While our companion mass upper limit was
ultimately unconstraining compared to the eclipse mass constraint, the
technique does show the power to measure relativistic effects with a
mission like {\it RXTE}.  If J1749 has another outburst (especially a longer
duration outburst which allows better coverage near eclipse), it may
be possible to place stronger constraints on the system masses.

\acknowledgments The authors thank the \rxte\ and \swift\
projects for support, and useful conversations with 
Diego Altamirano, 
Deepto Chakrabarty, and
Jean Swank.

\par {\it Facility:} \facility{RXTE (PCA)}

\clearpage

\begin{deluxetable}{lc}
\tablecaption{Orbital Parameters of SWIFT J1749\label{tab-orbit}}
\tablehead{\colhead{Parameter} & \colhead{Value}}
\startdata
\cutinhead{Pulsar Timing}
Solution Time Range (UTC)                              & 2007-04-14 to 2010-04-20\\
Barycentric pulse frequency, $f_o$ (Hz)       & 1035.840027850(65)\tablenotemark{a}\tablenotemark{b} \\
Pulsar frequency derivative, $|\dot f|$ (Hz s$^{-1}$)  & $<1.5\times 10^{-12}$ \tablenotemark{c} \\
Projected semimajor axis, $a_x \sin i$ (lt-s)          & 1.899494(12) \\
Binary orbital period, $P_b$ (s)                       & 31740.719(8) \\
Time of ascending node, $T_{\rm asc}$                  & 2455301.1522542(5) \tablenotemark{d} \\
Orbital eccentricity, $e$                              & 4.2(1.5)$\times 10^{-5}$\\
Pulsar mass function, $f_x$ ($10^{-2} M_\sun$)         & 5.463(8)\\
Maximum Power, $Z_{\rm max}^2$                         & 2362 \\
Timing residuals, $\sigma_{\rm toa}$ ($\mu$s)          & 92 \\
Solution quality, $\chi^2$ for $\nu$ d.o.f.            & 225 / 60 d.o.f. \\
\cutinhead{Eclipse Timing}
Eclipse duration, $T_{\rm ecl}$ (s)                    & 2172(13) \\
Egress duration, s                                     & 30(12) \\
Solution quality, $\chi^2$ for $\nu$ d.o.f.            & 372 / 372 d.o.f. \\
\cutinhead{Joint Solution\tablenotemark{e}}
Companion mass, $M_c$ ($M_\sun$)                       & 0.62--0.81 \\
Companion radius, $R_c$ ($M_\sun$)                     & 0.85--0.92 \\
Inclination, $i$ (degrees)                             & 76.65--77.93 \\
Right Ascension (J2000)\tablenotemark{f}               & 17$^{\rm h}$49$^{\rm m}$31\fs94\\
Declination (J2000)                                    & $-28\arcdeg08\arcmin05\farcs8$\\
\enddata
\tablenotetext{a}{Uncertainties are $1\sigma$ in the last quoted digits, and 
have been scaled by $\sqrt{\chi^2/\nu}$.}
\tablenotetext{b}{Pulse frequency of the pulsar first harmonic.}
\tablenotetext{c}{95\% upper limit.}
\tablenotetext{d}{Julian days, referred to TDB timescale.}
\tablenotetext{e}{Assuming neutron star mass, $M_x$, between 1.4--2.2 $M_\sun$.}
\tablenotetext{f}{Pulsar celestial position (Yang et al. 2010).}
\end{deluxetable}

\begin{deluxetable}{llcc}
\tablecaption{Eclipse Epochs of SWIFT J1749.4$-$2807\label{tab-eclipse}}
\tablehead{\colhead{No.} & \colhead{Time\tablenotemark{a}} & \colhead{Type} & Offset\tablenotemark{b}}
\startdata
 1 &    2455303.46026 & Egress &  \phs1098 $\pm$ \phn3 \\
 2 &    2455306.39913 & Egress &  \phs1091 $\pm$    10 \\
 3 &    2455307.47573 & Ingres &  $-$1113  $\pm$    12 \\
\enddata
\tablenotetext{a}{Julian day epoch of eclipse ingress/egress, referred to TDB.}
\tablenotetext{b}{Offset in seconds from time of nearest pulsar longitude = 90$^\circ$, $T90$.}
\end{deluxetable}

\clearpage

\clearpage

\clearpage

\clearpage

\end{document}